\documentclass[reprint, secnumarabic, amssymb, amsmath, aip,cha, groupedaddress,frontmatterverbose]{revtex4-1}
\usepackage[dvips]{graphicx}
\usepackage{bm}
\usepackage[dvips]{color}

\begin{document}
\title{Magnetic and electric behaviors of DyMn$_2$O$_5$: effect of hole doping}
\author{P. Dutta$^1$}
\email{prabirdutta1989@gmail.com}
\author{M. Das$^1$}
\author{S. Mukherjee$^1$}
\author{S. Chatterjee$^2$}
\author{S. Giri$^1$}
\author{S. Majumdar$^1$}
\email{sspsm2@iacs.res.in}
\affiliation{$^1$School of Physical Sciences, Indian Association for the Cultivation of Science, 2 A \& B Raja S. C. Mullick Road, Jadavpur, Kolkata 700 032, India}
\affiliation{$^2$UGC-DAE Consortium for Scientific Research, Kolkata Centre, Sector III, LB-8, Salt Lake, Kolkata 700 106, India}
\begin{abstract}
DyMn$_2$O$_5$ is an intriguing multiferroic material showing multiple magnetic, electric and structural transitions. We present here the systematic study on the effect of Sr doping at the Dy site of DyMn$_2$O$_5$ through magnetic and dielectric measurements. Doping of divalent Sr at the Dy site is expected to enhance the Mn$^{4+}$:Mn$^{3+}$ ratio and it will also dilute the Dy site. Our study indicates large enhancement in the magnetic anomaly observed close to 43 K, which we believe to be related to the increased ferromagnetic correlations on Sr doping. Gradual increase in coercive field at 3 K with the Sr doping and decrease in bond length of adjacent Mn$^{4+}$ ions further support the enhancement of ferromagnetic corelations in the system. The parent sample shows a large magnetocaloric effect around 12 K, the magnitude of which found to decrease with increasing Sr concentration. The doping also enhances the anomaly at around 28 K observed in the dielectric permittivity versus temperature data, and this anomaly was earlier claimed to be associated with the spin reorientation as well as a simultaneous transition from one ferroelectric state to other. The electric orderings observed below 25 K are found to be susceptible to the applied magnetic field, and supports the view of Ratcliff II {\it et al.}(Phys. Rev. B {\bf 72}, 060407(R)(2005)) of concomitant changes in the magnetic structure associated with the multiple electric transitions. 
\end{abstract}
\maketitle

\section{Introduction}
Orthorhombic RMn$_2$O$_5$ (R = rare-earth ion) type manganites have been studied extensively in recent times due to their exotic magnetic as well as multiferroic properties.~\cite{ABRAHAMS-dmojcp, Tokura-dmo, Hur-dmo, cruz-rmoprb, hur-nat, inomata-rmojpcm, nakamur-ferro, sumanta-nmo, yang-sr, sanina-jpcm, isao-ferro}  RMn$_2$O$_5$ is a mixed valent material consisting of both Mn$^{3+}$ and Mn$^{4+}$ ions in equal proportion. These materials show fascinating multiple magnetic transitions as a function of temperature. The magnetic interactions in these compounds are quite complicated with the presence of frustration.~\cite{cheong-nm} The spin induced electric polarization in these compounds primarily originates from exchange striction mechanism, as the magnetic structures are nearly colinear.~\cite{Jeroen-exstric,sumanta-rmoprl}

\begin{figure}[t]
\centering
\includegraphics[width = 8 cm]{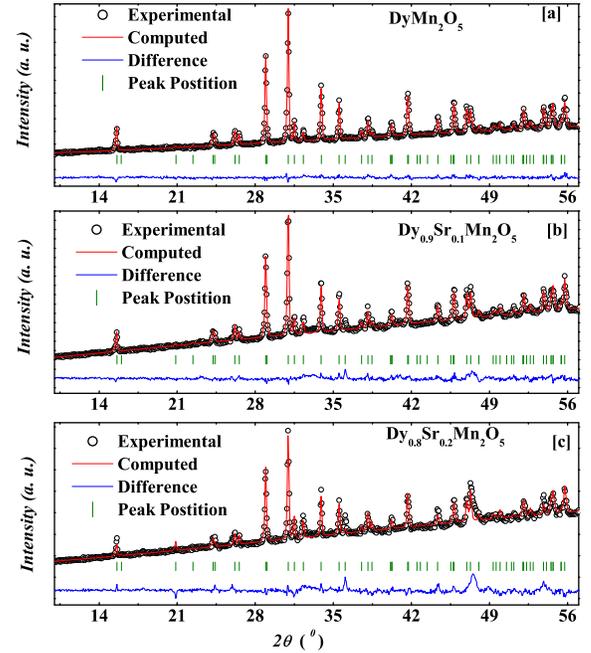}
\caption{Powder X-ray diffraction patterns of Dy$_{1-x}$Sr$_{x}$Mn$_{2}$O$_5$ ($x$ = 0.0, 0.1 and 0.2) along with Reitveld refinement patterns (red lines). The green ticks indicate the position of the Bragg peaks, while the blue line represent the difference between experimental data and the fitted curves.} 
\label{xrd}
\end{figure}
\par
The crystal structure of RMn$_2$O$_5$ consists of edge-sharing Mn$^{4+}$O$_6$ octahedra and Mn$^{3+}$O$_5$ pyramids. The Mn$^{4+}$O$_6$ octahedra form ribbons running along the $c$ axis. Two such adjacent ribbons are separated alternatively by layers of Mn$^{3+}$ and R$^{3+}$ ions. The nearest and next nearest neighbour magnetic interactions $J_1$ and $J_2$ are between two Mn$^{4+}$ ions through R$^{3+}$ and Mn$^{3+}$ layers respectively. $J_2$ is found to be stronger and it is ferromagnetic (FM) in nature.  When viewed along the $c$ direction, five Mn ions (two Mn$^{4+}$ and three Mn$^{3+}$) are found to form loop with the R$^{3+}$ ion at the centre. The interactions $J_3$ and $J_4$ between neighbouring Mn$^{4+}$ and Mn$^{3+}$ ions in the loop are antiferromagnetic (AFM) in nature. Similarly $J_5$ between Mn$^{3+}$ and Mn$^{3+}$ is also AFM. Evidently, due to the presence of AFM interactions ($J_3$, $J_4$ and $J_5$) in the loop, a certain degree of magnetic frustration is obvious.~\cite{Noda-rmo, radaelli-jpcm}

\begin{figure*}[t]
\centering
\includegraphics[width = 14 cm]{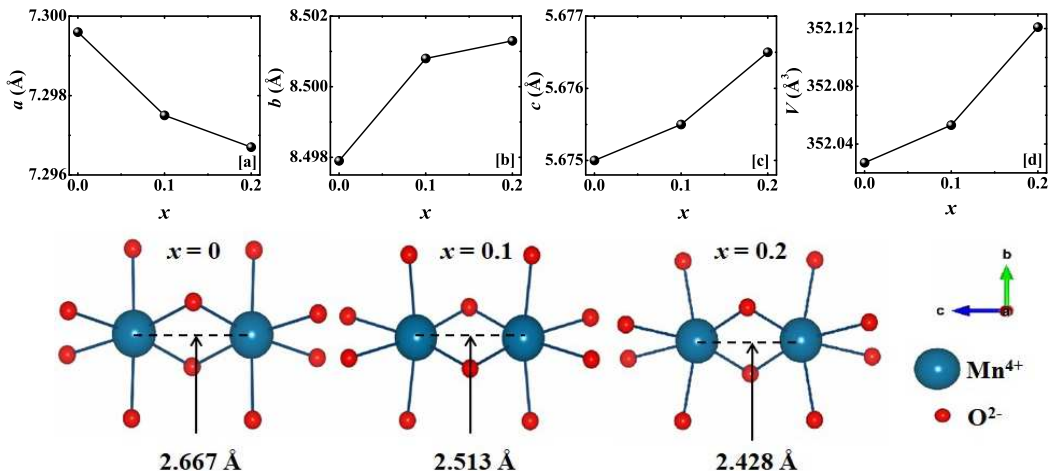}
\caption{The upper panel shows the variation of lattice parameters as a function of Sr concentration. The lower panel shows a perspective view of the bond length between two neighboring Mn$^{4+}$ ions on the chain along the $c$ axis. The bond distance decreases with increasing Sr concentration.} 
\label{structure}
\end{figure*}
\par
In the present work, we studied the compound DyMn$_2$O$_5$ and its hole doped derivatives. The parent compound orders magnetically from a paramagnetic (PM) to an incommensurate antiferromagnetic (IC-AFM)  state below  43 K. Powder neutron diffraction studies, both on single crystalline  and polycrystalline samples, indicate that this IC-AFM phase continues to exist down to 9 K.~\cite{sumanta-sr, Blake-prb}  Below 9 K, the magnetic structure becomes commensurate (C-AFM) in nature with the concomitant ordering of the Dy sublattice. A small fraction of the IC-AFM phase still remains untransformed even  below 9 K. However, there are several other reports claiming the  presence of additional magnetic transitions intermediate between 9 and 43 K. ~\cite{Zhao-sr} Ratcliff II {\it et al} finds a significantly complex phase diagram  with other magnetic transitions at 40 K, 28 K, 20 K based on their single-crystal neutron diffraction studies, along with the coexistence of multiple magnetic phases (both C-AFM and IC-AFM) below 28 K.~\cite{Ratcliff-prb}

\begin{figure}[t]
\centering
\includegraphics[width = 8 cm]{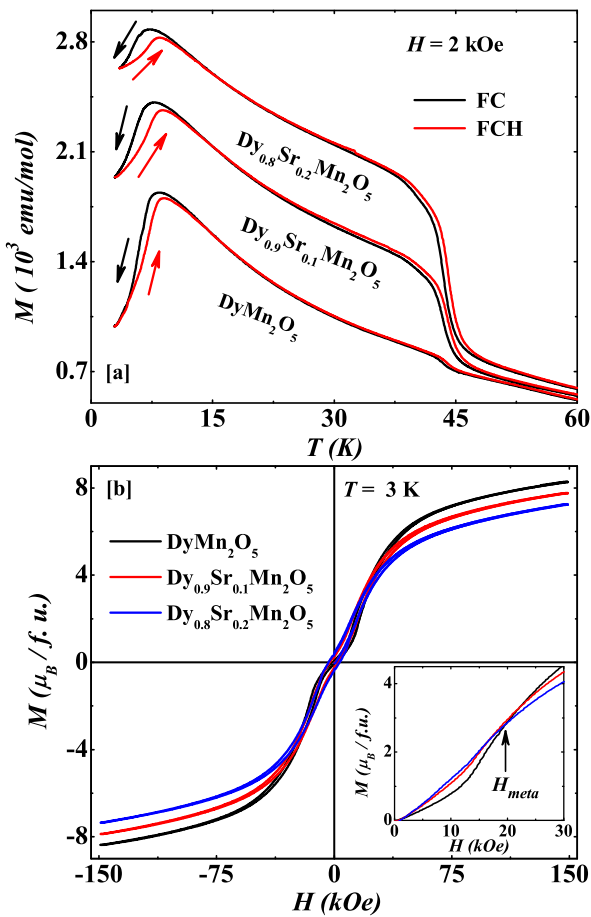}
\caption{(a) shows the thermal variation of magnetization in presence of 2 kOe of applied field for Dy$_{1-x}$Sr$_{x}$Mn$_{2}$O$_5$ ($x$ = 0.0, 0.1 and 0.2). (b) represents the isothermal magnetization data recorded at 3 K for the above three samples.} 
\label{mag}
\end{figure}

\par
Just below the magnetic transition at 43 K, the DyMn$_2$O$_5$ shows a transformation into a ferroelectric (FE) state with the emergence of spontaneous polarization at 40 K (from paraelectric to FE1 state). It also shows multiple anomalies in polarization, some of which are linked with the spin ordering temperatures. Polarisation shows anomalies due to the change in the ferroelectric states at 28 K (from FE1 to FE2) and 20 K (from FE2 to FE3) while cooling. Finally, it shows clear anomaly at 9 K due to the onset of a ferrielectric state, and it is linked to the magnetic ordering of rare-earth sublattice.~\cite{Tokura-dmo, Zhao-sr, Sushkov-prb}   

\par
DyMn$_2$O$_5$ is a mixed-valent manganite containing both Mn$^{3+}$ and Mn$^{4+}$ ions. The fascinating and multiple phase transitions, as well as multiferroic properties, is connected to the mixed valent nature of Mn. In the present work, we have tuned the Mn$^{3+}$ and  Mn$^{4+}$ ratio by doping divalent Sr at the Dy site. The present work aims to ascertain the role of enhanced Mn$^{4+}$ concentration on the magnetic and multiferroic properties of this complex oxide.  

\section{Experimental Details}
The desired polycrystalline samples of nominal compositions Dy$_{1-x}$Sr$_{x}$Mn$_{2}$O$_5$ ($x$ = 0.0, 0.1 and 0.2) were prepared by the conventional solid-state reaction method. Required amount of Dy$_2$O$_3$,  SrCO$_3$, and Mn(CH$_3$COO)$_2$.4H$_2$O  with purity $\geq$ 99.9\% were mixed thoroughly and heated at 900$^{\circ}$C for 24 h. The resultant products were sintered at 1150$^{\circ}$C for 72 h in pellet form with three intermediate grindings. Room temperature powder X-ray diffraction (PXRD) patterns were recorded in a PANalytical X-ray diffractometer using Cu K$\alpha$ radiation. The dc magnetization ($M$) was measured using a vibrating sample magnetometer attached with the commercial cryogen-free high magnetic field system from Cryogenic Ltd., U.K. in the temperature ($T$) range between 3-300 K in the presence of applied magnetic field ($H$) = 0-150 kOe. Dielectric permittivity ($\epsilon$) was measured by an LCR meter (Agilent E4980a) attached to a closed cycle refrigerator.  A thin rectangular and polished piece of each sample was used for $\epsilon$ measurement, and electrodes are made on two polished surfaces using thin layers of silver paste. Pyroelectric current density ($J_P$) was recorded in an electrometer (Keithley,
model 6517B) connected to the PPMS-II system (Quantum Design). First the sample was cooled down to 100 K and then a poling electric filed ($E_{pole}$) of +5 kV/cm (or -5 kV/cm) was applied to the sample until the sample was further cooled down to 2 K. At 2 K the sample was short-circuited for a time of 40 min to release the accumulated charges in the sample. Finally the sample was heated at a constant temperature sweep rate of 5 K/min and the $J_P$ data were collected during this warming process. $J_P$ was also measured in presence of 50 kOe magnetic field in two different cooling conditions namely (i) zero field cooled and (ii) field cooled. 

\begin{figure*}[t]
\centering
\includegraphics[width = 14 cm]{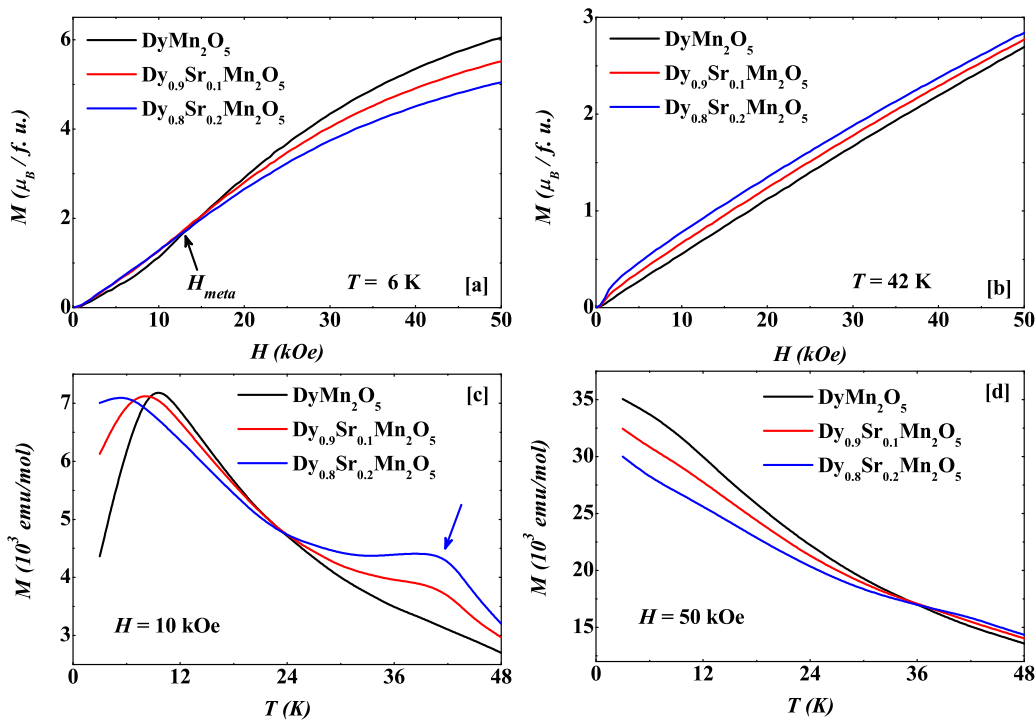}
\caption{(a) and (b) show the isothermal magnetization as a function of applied field at two different constant temperatures for Dy$_{1-x}$Sr$_{x}$Mn$_{2}$O$_5$ ($x$ = 0.0, 0.1 and 0.2) samples. (c) and (d) represent magnetization versus temperature data in presence of two different magnetic field for the above three samples.} 
\label{isotherms}
\end{figure*}

\section{Results}
Fig.~\ref{xrd} shows the PXRD patterns of the studied compounds measured at room temperature. We have analyzed our PXRD data using MAUD software package ~\cite{maud-jap}, and the pattern can be well fitted considering orthorhombic structure with space group $Pbam$. The variation of lattice parameters ($a$, $b$, and $c$), and lattice volume ($V$) with $x$ is depicted in the upper panel of fig.~\ref{structure}. The parameters $b$ and $c$ increase with $x$, while $a$ decreases with an overall increase in the lattice volume. This is because the 2+ ionic radius of Sr is larger than 3+ ionic radius of Dy. Fig.~\ref{structure} (bottom panel) shows the variation of  bond distance  between adjacent Mn$^{4+}$ ions on the chain, and a clear decrease in the bond length is observed with $x$.

\par
Fig.~\ref{mag} (a) denotes the $M$ versus $T$ data recorded at  $H$ = 2 kOe. The doped samples show two prominent features in the magnetic data, namely at around 43 K and 9 K. The 43 K anomaly can be assigned to the long-range magnetic ordering into an IC-AFM state. On the other hand, the 9 K anomaly is due to the magnetic ordering of Dy sublattice.  Although the 9 K anomaly is quite strong in the pure sample, the 43 K one is rather weak, and it is visible only as a small kink. It is evident that the strength of the 43 K anomaly turns considerably stronger in the doped compositions, while the 9 K anomaly turns weaker. One notable observation is the presence of thermal hysteresis both around the 9 K and 43 K transitions. The 9 K thermal hysteresis is quite clear, with the width of the hysteresis being more than 2 K for $x$ = 0.2 sample, while the hysteresis at 43 K transition has a width of 0.6 K for the same sample. The first order nature of the transition at 9 K was already known,~\cite{Ratcliff-prb, Mihailova-rmo} although the existing literature does not discuss on the order of the transition at 43 K. The width of the thermal hysteresis observed in the present work around 43 K is small, but it is reproducible in all three samples. The FE transition at 40 K as well as the magnetic transition at 43 are associated with lattice distortion. Therefore, the thermal hysteresis around 43 K is likely to be an effect of first order lattice distortion in the said temperature range.~\cite{Blake-prb, cruz-rmojap} 

\par
The magnetic susceptibility ($\chi = M/H$) versus $T$ data obey Curie-Weiss law above 120 K. We have obtained the  effective paramagnetic moment ($p_{eff}$) and Curie temperature (${\theta}_p$) from the linear 1/$\chi$ versus $T$ plot between 120-300 K and their values are tabulated in table 1. For the pure compound, the value of $p_{eff}$ is found to be 12.42 $\mu_B$/f.u., which is quite close to the value of 12.35 $\mu_B$/f.u. expected for the system with Mn$^{3+}$ in the high spin state. The negative value of ${\theta}_p$ indicates predominantly AFM correlations in the system. With Sr doping, the value of $p_{eff}$ gradually decreases. In table 1, we have compared the theoretical and experimental value of the moments in three compositions, and the experimentally observed moments closely follow the theoretical moment values. Here theoretical moment is calculated by the relation, $$p_{eff}^2 (theo) = n_1p_1^2 + n_2p_2^2 + n_3p_3^2,$$ where $n_1$, $n_2$, $n_3$ are respectively  the number of atoms of Dy, Mn$^{3+}$ and Mn$^{4+}$ per formula unit, and  $p_1$, $p_2$ and $p_3$ are their respective moments.  

\begin{figure}[t]
\centering
\includegraphics[width = 8 cm]{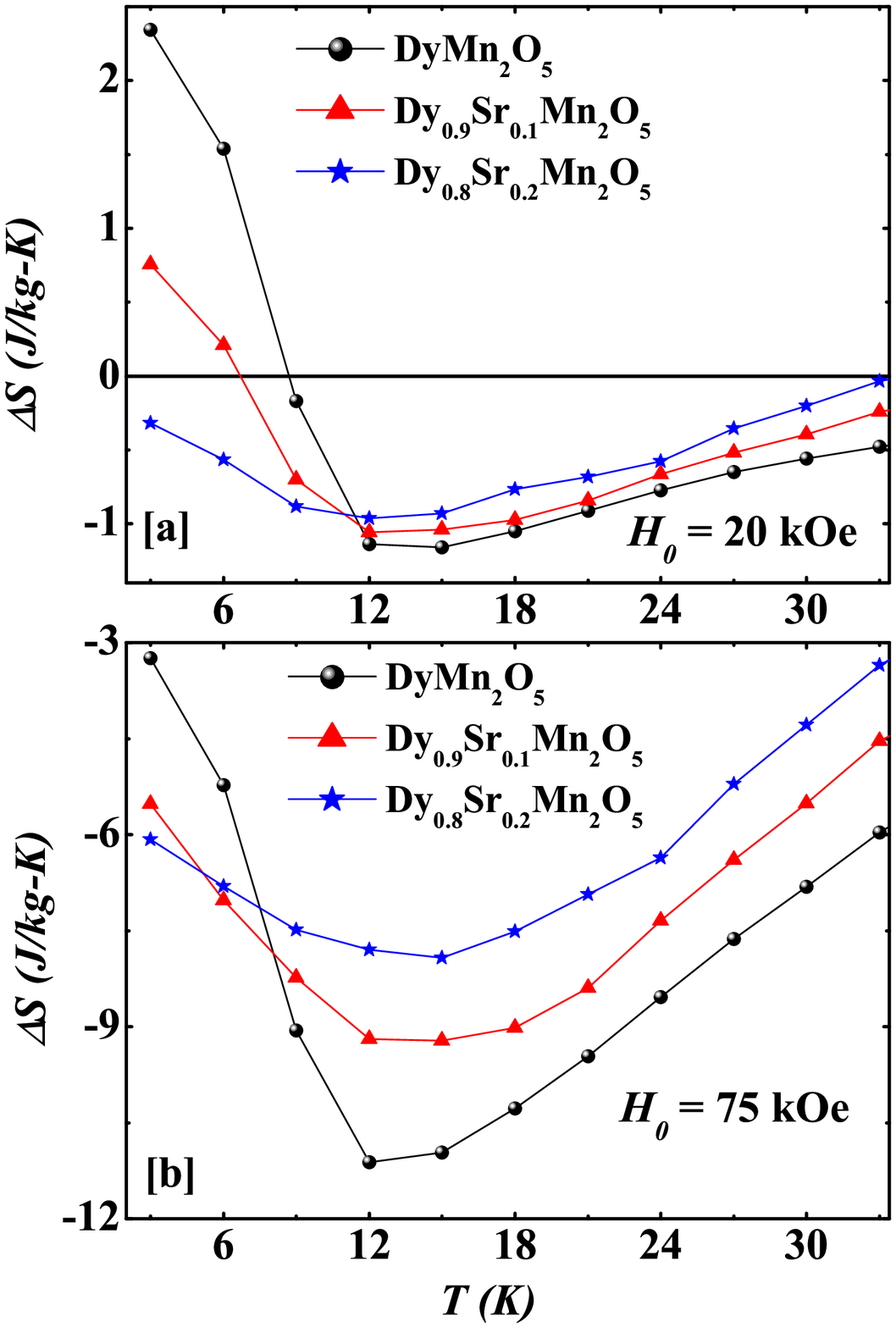}
\caption{(a) and (b)  show the thermal variation of $\Delta S$ for all three investigated samples with $H_0$ = 20 and 75 kOe respectively.} 
\label{mce}
\end{figure}

Fig.~\ref{mag} (b) shows the $M$ versus $H$ isotherms recorded at 3 K for the studied  samples. For the pure sample, an anomaly is observed around 20 kOe, which slowly disappears for the doped samples [see inset of fig.~\ref{mag} (b)]. The value of $M$ at the maximum applied $H$ ( = 150 kOe) decreases systematically with increasing Sr concentration, which is clear from the values recorded in table 1. However, none of the samples shows full saturation even at a field as high as 150 kOe. The pure sample shows a coercive field of $H_{coer}$ = 1.49 kOe at 3 K, which also gradually increases with increasing $x$ (see table 1). 

\par
To ascertain the magnetic nature of the pure and the doped samples, we compared  magnetic isotherms of all three samples at 42 K and 6 K, which are respectively shown in figs.~\ref{isotherms} (a) and (b). In the 6 K, isotherm, the metamagnetic transition occurs at a slightly lower field of  15 kOe, where $M-H$ curves of all three samples intersect with each other. $M$ is higher for the doped samples below 15 kOe, while pure sample attains a higher value of $M$ beyond 15 kOe. At 42 K (just below the PM to IC-AFM transition), the scenario is entirely different. Here $M$ is found to be higher in case of doped samples between 0 to 50 kOe of fields. 

\par
We have plotted the thermal variation of the magnetization of three samples recorded under the application of $H$ = 10 and 50 kOe [figs.~\ref{isotherms} (c) and (d)]. At 10 kOe, the PM to IC-AFM transition near 43 K becomes prominent with increasing Sr concentration. Below about 24 K, magnitude of $M$ of the pure sample turns larger than the doped ones. Presumably, it is related to the development of Dy$^{3+}$-Mn$^{4+}$ interaction in this range of $T$, and the doped samples with lower Dy concentration shows the lower value of $M$. One important observation from fig.~\ref{isotherms} (c) is that the peak around 9 K associated with the ordering of Dy sublattice shifts to lower $T$ with Sr substitution. The $M$ versus $T$ curves at 50 kOe [fig.~\ref{isotherms} (d)] do not show any peak like feature around 43 K or 9 K transition points. The values of  $M$ of the Sr-rich samples are found to be higher at higher temperatures (above 36 K), while they turn lower below the cross over point of 36 K.         
\begin{table}
\begin{tabular}{p{3cm} p{1.8cm} p{1.8cm} p{1.8cm}}
                    \hline \hline 
Parameters &  $x$ = 0.0 & $x$ = 0.1 & $x$ = 0.2 \\
                            \hline 
                                  
$p_{eff}$($\mu_B$/f.u.)(theo)&        12.35&     11.83&     11.31     \\ 

$p_{eff}$($\mu_B$/f.u.)(exp)&       12.42&     12.1&     11.7 \\
    
$\theta_p$ (K) &        $-$23&     $-$17&     $-$5     \\ 
 
$M_{15T}$ ($\mu_B$/f.u.)&        8.3&     7.8&     7.2     \\
  
$H_{coer}$ (kOe)&        1.49&     2.91&     3.55    \\

$\Delta S$ (Jkg$^{-1}$K$^{-1}$)& -11.2/-6.7 & -9.2/-5.8 & -7.9/-4.9 \\
($H_0$ = 75/50 kOe)\\   

\hline 
 \end{tabular}
\caption{Variation of effective paramagnetic moment ($p_{eff}$), Curie temperature ($\theta_p$), magnetic moment at 3 K under $H$ = 150 kOe ($M_{15T}$), coercive field at 3 K ($H_{coer}$) and magnetic entropy change ($\Delta S$) with Sr doping at the Dy site.}
\label{ap8}
\end{table} 
\begin{figure}[t]
\centering
\includegraphics[width = 8 cm]{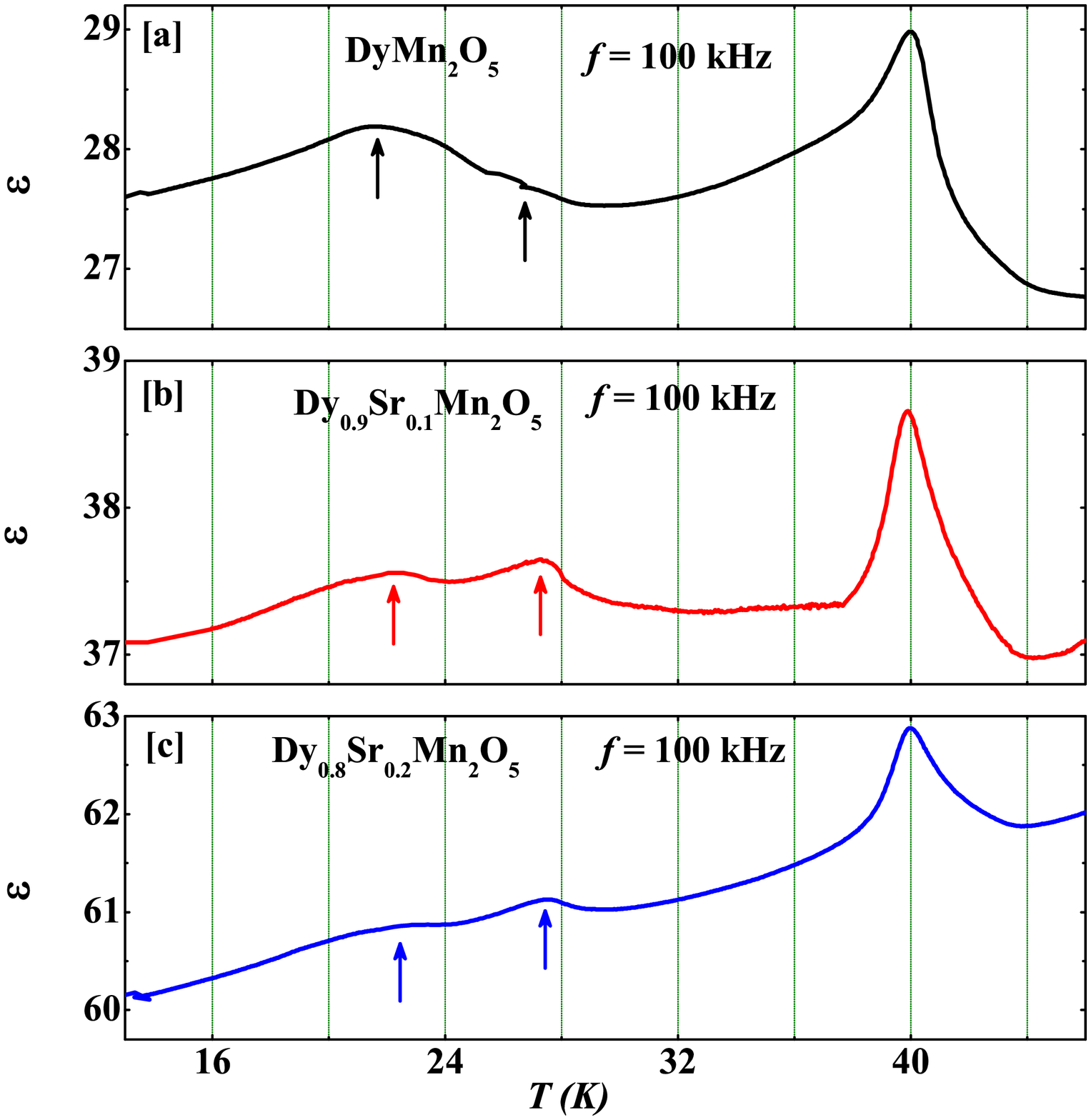}
\caption{(a), (b) and (c) show electric permittivity as a function of temperature for Dy$_{1-x}$Sr$_{x}$Mn$_{2}$O$_5$ ($x$ = 0.0, 0.1 and 0.2) samples measured with input ac voltage of frequency 100 kHz.} 
\label{dielectric}
\end{figure}

\par
Observing the sharp first-order phase transition around 9 K in the $M(T)$ data and the existence of metamagnetism, we have investigated the magnetocaloric effect (MCE) of this system. Change in the magnetic entropy ($\Delta S$) with the application of $H$ is a measure of MCE. One can easily calculate it from magnetization data by using Maxwell's relation 

\begin{equation}
{\Delta}S(0{\rightarrow}H_0)=\int^{H_0}_0\left(\frac{{\partial}M}{{\partial}T}\right)_HdH
\label{maxwell}
\end{equation} 

where ${\Delta}S(0{\rightarrow}H_0)$ denotes the entropy change for the change in $H$ from $0$ to $H_0$.~\cite{gs-mce} For this calculation, we have recorded isothermal $M(H)$ data between 3 K to 40 K with 3 K interval [not shown in here] . Before every $M(H)$ measurement, it was confirmed that the sample was in a thermally demagnetized state. This state was achieved by cooling the sample from PM region to 3 K and then heated back to the respective temperature of measurements in the absence of magnetic field. The same procedure was followed for the other two samples. Variation of $\Delta S$ with $T$ for $H_0$ = 20 and 75 kOe are plotted in figs.~\ref{mce} (a) and (b) respectively. At 20 kOe, $x$ = 0.0 and 0.1 samples show both conventional and  inverse MCE ($\Delta S <$ 0 and $>$ 0 respectively). The MCE is conventional down to 9 K and 7 K respectively for these two samples, below which $\Delta S$ turns positive. This change of sign in $\Delta S$ can be related to metamagnetism observed at low temperature (below 9 K). The $x$ = 0.2 sample shows a low-$T$ rise in $\Delta S$ like other members, but it remains negative down to the lowest-$T$ of measurement.  

\begin{figure}[t]
\centering
\includegraphics[width = 8 cm]{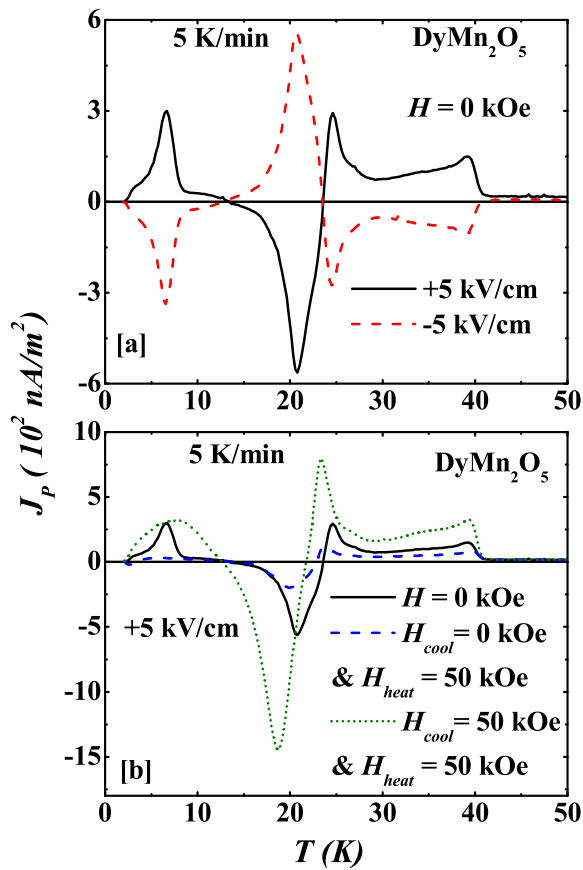}
\caption{(a) represents the thermal variation of pyroelectric current density ($J_P$) measured at two different poling electric fields (+5 kV/cm and -5 kV/cm) in absence of magnetic field for the DyMn$_{2}$O$_5$ sample. $J_P(T)$ data recorded in presence of 50 kOe magnetic field in two different cooling conditions namely (i) zero field cooled (blue curve) and (ii) field cooled (green curve) are depicted in (b). During each measurement the warming rate is 5 K/min.} 
\label{pyro}
\end{figure}

\par
The parent sample shows a large value of $\Delta S$ at 75 kOe with a peak  of magnitude 11.2 Jkg$^{-1}$K$^{-1}$ at  12 K.  At $H_0$ = 50 kOe, the MCE is slightly less, but still has a significant value (6.7 Jkg$^{-1}$K$^{-1}$). The magnitude of MCE shows systematic decrement  with increasing Sr concentration (see table 1).  The MCE is found to be conventional ($\Delta S <$ 0) for all three samples at 75 kOe. We have calculated relative cooling power of the sample, which is defined as RCP = $\Delta S_{max} \times  \delta T_{fwhm}$, where $\Delta S_{max}$ and $\delta T_{fwhm}$  are  the peak value and peak width (full width at half maximum) of the $\Delta S(T)$ curve. ~\cite{ali-rcp,hernado-rcp}  RCP is one of the important parameters for practical application in a refrigeration system. The calculated RCP is found to be 316 J/kg for the pure sample for $H_0$ = 75 kOe,  which is comparable or higher than many other magnetocaloric materials among transition metal oxides.~\cite{phan-jmmm}  
    
\par
Since the parent sample is a well known multiferroic material, we have also addressed the impact of Sr doping in dielectric properties. $\epsilon$ was recorded as a function of $T$ at 100 kHz frequency for all the three samples and the data recorded between  14-46 K (heating leg) are shown in fig.~\ref{dielectric}). All three samples show a prominent peak at 40 K in $\epsilon(T)$, which is associated with the ferroelectricity driven by IC-AFM to C-AFM magnetic transition.  The parent sample shows a weak kink like feature at around 27 K, which can be correlated with the inter-ferroelectric phase transition (FE1 to FE2).~\cite{Tokura-dmo, Zhao-sr, Sushkov-prb}

\par
Interestingly, 27 K transition becomes much pronounced in the Sr-doped samples. The third anomaly in $\epsilon(T)$ is observed in the form of a broad hump around 21 K, which signifies a transition to the third ferroelectric phase (FE3). The 21 K feature is strong in the pure sample and turns weaker on Sr doping.

\par
In order to ascertain the electric order and its correlation with the applied magnetic field, we have recorded $J_P$ as a function of $T$ in different protocols for the parent sample (as shown in fig.~\ref{pyro}). In addition to the poling with an external electric field, the sample was field-cooled in magnetic field. Fig.~\ref{pyro} (a) shows the variation of $J_P$ of parent sample with $T$ in absence of any external $H$.  Reversal of $J_P(T)$ with the sign of $E_{pole}$ suggests  the existence of electric order (ferroelectric/ferrielectric)  in the sample. Our $J_P(T)$ curves show number of anomalies at different transition temperatures which match well with the $\epsilon(T)$ data as well as the previously published results~\cite{Zhao-sr}. The anomalies observed around 40 K, 25 K and 21 K (see fig.~\ref{pyro}a) indicate paraelectric to FE1, FE1 to FE2, and FE2 to FE3 transition respectively. Whereas, peak around 7 K is related to the Dy$^{3+}$ ordering temperature. Fig. ~\ref{pyro} (b) shows the same $J_P(T)$ of $x$ = 0 sample under external $H$ (under magnetic field-cooling  and Zero-field-cooling). Interestingly, the 7 K peak completely vanishes when we have measured the $J_P(T)$ in presence of 50 kOe magnetic field after being cooled in zero magnetic field. On the other hand, a broad hump like behaviour is observed around 7 K in magnetic field cooled heating data. This is a clear indication of the strong magneto-electric effect  associated with the Dy ordering. One additional observation of field cooling is the large enhancement of  the magnitude of $J_P$ around the peaks at 21 and 25 K. The 21 K and 25 K transition temperature are shifted to lower $T$, whereas the 40 K transition  remains almost unchanged under $H$. This indicates that the PE to FE1 transition is less susceptible to an external magnetic field. It is evident that $J_P$ vanishes at the lowest $T$, which is connected to the ferrielectric nature of the samples. We refrained from calculating polarization by integrating  our $J_P(T)$ data, as the technique may  be inapplicable to the present ferri-electric materials.~\cite{Tokura-dmo} 

\section{Summary and conclusions}
We observe several changes in magnetic and dielectric properties on doping Sr at the Dy site of the exotic multiferroic material, DyMn$_2$O$_5$. The most noticeable changes on Sr doping are: (i) there is a  large enhancement of the magnetic anomaly at the paramagnetic to IC-AFM transition around 43 K, and (ii) the coercive field of the sample recorded at 3 K increases systematically. These indicate that Sr substitution enhances the FM correlation in the system.  As already mentioned, Mn$^{4+}$O$_6$ octahedra form ribbons running along the $c$ axis, and neutron diffraction data indicate that the $c$-component of the Mn$^{4+}$ moments are aligned ferromagnetically. In the $a-b$ plane, Mn$^{4+}$ ions are separated by Mn$^{3+}$ ions, and the interaction is AFM. On replacing Dy by divalent Sr, the fraction of  Mn$^{4+}$ ions gets increased. Since there exists an FM component in Mn$^{4+}$-Mn$^{4+}$ interaction, it can lead to the increase of the overall ordered moment, which gets reflected at the 43 K transition. It is  to be noted that Mn$^{4+}$-Mn$^{4+}$ direct exchange also favors an FM alignment. On doping Sr at the Dy site, the Mn$^{4+}$-Mn$^{4+}$ bond distance decreases (the change is about 9\%, and it is depicted in fig.~\ref{structure}), paving the path for stronger direct exchange and consequently enhanced FM moment. The same argument  explains the increased coercive field in Sr-doped samples.               

\par
Contrary to the change observed around 43 K, the transition at 9 K becomes weaker with Sr doping. This is understandable as Dy sublattice gets diluted with Sr substitution. The metamagnetic transition observed below the ordering of Dy-sublattice, also gets weaker with Sr substitution due to the same mechanism. If one compares the isothermal $M$ versus $H$ behavior of the studied samples at 3 K, $M$ is found to be higher in the maximum doped sample at the lower field (precisely below the metamagnetic transition field $H_{meta}$), while $M$ for the pure sample dominates above $H_{meta}$. This is related to the effect of two factors, namely, (i) enhanced FM component and (ii) weaker metamagnetism in doped samples. At lower fields ($<H_{meta}$)  $M$ is large in doped samples due to the excess FM component originating from Mn$^{4+}$ ions. While beyond $H > H_{meta}$, stronger FM component in the pure sample from Dy sublattice (the FM component originates from the alignment of Dy moments by $H$) makes $M$ to be lower in Sr-doped compositions.    

\par
One crucial observation is the occurrence of large MCE peaking around 12 K for all the samples. At low values of a field change (0 to 20 kOe), the pure and 10\% doped samples show a positive value of $\Delta S$ (inverse MCE) below 9 and 7 K respectively. Such inverse MCE is clearly related to the C-AFM phase of the Dy-sublattice, which disappears at higher fields due to metamagnetic transition (see fig. 4 (b)). The large conventional MCE (negative $\Delta S$) at higher fields (beyond 20 kOe) is clearly related to the metamagnetic transition of the Dy sublattice. It reduces with the Sr doping due to the dilution of the rare-earth moment. 

\par
The dielectric anomaly observed in DyMn$_2$O$_5$ also get influenced by the Sr doping. The most important changes are found  between 20 to 30 K. Previous literature claimed that DyMn$_2$O$_5$ undergoes FE1 to FE2 transition at 28 K, and there is also report of a change in the magnetic structure at the same temperature. For the pure sample, we see a weak feature in our $\epsilon$ versus $T$ data around 27 K and it is consistent with the feature reported by other groups on polycrystalline sample.~\cite{Zhao-sr} The pure sample also shows another feature just above 21 K, which was earlier found to be related to FE2 to FE3 transition. On Sr doping, we find that the feature at 20 K turns weaker, and a strong peak arises around 28 K. It is possible that Mn$^{4+}$ favors the FE2 phase, resulting an enhanced anomaly at 27 K. ~\cite{Zhao-sr, Zhao-Al-doped, Zhao-Y-doped} 

\par
The pyroelectric current measurement provide us with some important insight on the interesting material. Firstly, application of $H$ = 50  on the zero-field-cooled state destroys the peak in $J_P$ at 7 K. This is possibly connected to the metamagnetic transition observed in the sample around 20 kOe. The resulting magnetic structure may antagonize the ferri-electric order at low temperature. Secondly, the peaks observed in $J_P$ between 20 and 25 K corresponding to FE2 to FE3 and FE2 to FE1 transitions get strongly modified with the applied $H$. This supports the view of Ratcliff II {\it et al.}~\cite{Ratcliff-prb}, who observed simultaneous changes in magnetic structure at those multiple FE transitions.    

\par
In conclusion, we have successfully doped Sr at the Dy site of DyMn$_2$O$_5$. The magnetic and dielectric properties show some systematic changes with the Sr concentration. We attribute these changes to the dilution of Dy sublattice as well as enhanced FM interaction due to the increase in Mn$^{4+}$ ions in the system.

\section{Acknowledgment}
P. Dutta would like to thank Science and Engineering Research Board, India for his national postdoctoral fellowship (project number PDF/2017/001061). M. Das thanks CSIR, India for her research fellowship. This work is supported by the financial grant from DST-SERB project (EMR/2017/001058). Authors would also like to thank Department of Science and Technology, India for low temperature high magnetic field facilities at UGC-DAE Consortium for Scientific Research, Kolkata Centre. 


\begin{thebibliography}{30}%
\makeatletter
\providecommand \@ifxundefined [1]{%
 \@ifx{#1\undefined}
}%
\providecommand \@ifnum [1]{%
 \ifnum #1\expandafter \@firstoftwo
 \else \expandafter \@secondoftwo
 \fi
}%
\providecommand \@ifx [1]{%
 \ifx #1\expandafter \@firstoftwo
 \else \expandafter \@secondoftwo
 \fi
}%
\providecommand \natexlab [1]{#1}%
\providecommand \enquote  [1]{``#1''}%
\providecommand \bibnamefont  [1]{#1}%
\providecommand \bibfnamefont [1]{#1}%
\providecommand \citenamefont [1]{#1}%
\providecommand \href@noop [0]{\@secondoftwo}%
\providecommand \href [0]{\begingroup \@sanitize@url \@href}%
\providecommand \@href[1]{\@@startlink{#1}\@@href}%
\providecommand \@@href[1]{\endgroup#1\@@endlink}%
\providecommand \@sanitize@url [0]{\catcode `\\12\catcode `\$12\catcode
  `\&12\catcode `\#12\catcode `\^12\catcode `\_12\catcode `\%12\relax}%
\providecommand \@@startlink[1]{}%
\providecommand \@@endlink[0]{}%
\providecommand \url  [0]{\begingroup\@sanitize@url \@url }%
\providecommand \@url [1]{\endgroup\@href {#1}{\urlprefix }}%
\providecommand \urlprefix  [0]{URL }%
\providecommand \Eprint [0]{\href }%
\providecommand \doibase [0]{http://dx.doi.org/}%
\providecommand \selectlanguage [0]{\@gobble}%
\providecommand \bibinfo  [0]{\@secondoftwo}%
\providecommand \bibfield  [0]{\@secondoftwo}%
\providecommand \translation [1]{[#1]}%
\providecommand \BibitemOpen [0]{}%
\providecommand \bibitemStop [0]{}%
\providecommand \bibitemNoStop [0]{.\EOS\space}%
\providecommand \EOS [0]{\spacefactor3000\relax}%
\providecommand \BibitemShut  [1]{\csname bibitem#1\endcsname}%
\let\auto@bib@innerbib\@empty
\bibitem [{\citenamefont {Abrahams}\ and\ \citenamefont
  {Bernstein}(1967)}]{ABRAHAMS-dmojcp}%
  \BibitemOpen
  \bibfield  {author} {\bibinfo {author} {\bibfnamefont {S.~C.}\ \bibnamefont
  {Abrahams}}\ and\ \bibinfo {author} {\bibfnamefont {J.~L.}\ \bibnamefont
  {Bernstein}},\ }\href@noop {} {\bibfield  {journal} {\bibinfo  {journal} {J.
  Chem. Phys.}\ }\textbf {\bibinfo {volume} {46}},\ \bibinfo {pages} {3776}
  (\bibinfo {year} {1967})}\BibitemShut {NoStop}%
\bibitem [{\citenamefont {Higashiyama}\ \emph {et~al.}(2004)\citenamefont
  {Higashiyama}, \citenamefont {Miyasaka}, \citenamefont {Kida}, \citenamefont
  {Arima},\ and\ \citenamefont {Tokura}}]{Tokura-dmo}%
  \BibitemOpen
  \bibfield  {author} {\bibinfo {author} {\bibfnamefont {D.}~\bibnamefont
  {Higashiyama}}, \bibinfo {author} {\bibfnamefont {S.}~\bibnamefont
  {Miyasaka}}, \bibinfo {author} {\bibfnamefont {N.}~\bibnamefont {Kida}},
  \bibinfo {author} {\bibfnamefont {T.}~\bibnamefont {Arima}}, \ and\ \bibinfo
  {author} {\bibfnamefont {Y.}~\bibnamefont {Tokura}},\ }\href@noop {}
  {\bibfield  {journal} {\bibinfo  {journal} {Phys. Rev. B}\ }\textbf {\bibinfo
  {volume} {70}},\ \bibinfo {pages} {174405} (\bibinfo {year}
  {2004})}\BibitemShut {NoStop}%
\bibitem [{\citenamefont {Hur}\ \emph {et~al.}(2004{\natexlab{a}})\citenamefont
  {Hur}, \citenamefont {Park}, \citenamefont {Sharma}, \citenamefont {Guha},\
  and\ \citenamefont {Cheong}}]{Hur-dmo}%
  \BibitemOpen
  \bibfield  {author} {\bibinfo {author} {\bibfnamefont {N.}~\bibnamefont
  {Hur}}, \bibinfo {author} {\bibfnamefont {S.}~\bibnamefont {Park}}, \bibinfo
  {author} {\bibfnamefont {P.~A.}\ \bibnamefont {Sharma}}, \bibinfo {author}
  {\bibfnamefont {S.}~\bibnamefont {Guha}}, \ and\ \bibinfo {author}
  {\bibfnamefont {S.-W.}\ \bibnamefont {Cheong}},\ }\href@noop {} {\bibfield
  {journal} {\bibinfo  {journal} {Phys. Rev. Lett.}\ }\textbf {\bibinfo
  {volume} {93}},\ \bibinfo {pages} {107207} (\bibinfo {year}
  {2004}{\natexlab{a}})}\BibitemShut {NoStop}%
\bibitem [{\citenamefont {Cruz}\ \emph
  {et~al.}(2006{\natexlab{a}})\citenamefont {Cruz}, \citenamefont {Yen},
  \citenamefont {Lorenz}, \citenamefont {Gospodinov}, \citenamefont {Chu},
  \citenamefont {Ratcliff}, \citenamefont {Lynn}, \citenamefont {Park},\ and\
  \citenamefont {Cheong}}]{cruz-rmoprb}%
  \BibitemOpen
  \bibfield  {author} {\bibinfo {author} {\bibfnamefont {C.~R.~d.}\
  \bibnamefont {Cruz}}, \bibinfo {author} {\bibfnamefont {F.}~\bibnamefont
  {Yen}}, \bibinfo {author} {\bibfnamefont {B.}~\bibnamefont {Lorenz}},
  \bibinfo {author} {\bibfnamefont {M.~M.}\ \bibnamefont {Gospodinov}},
  \bibinfo {author} {\bibfnamefont {C.~W.}\ \bibnamefont {Chu}}, \bibinfo
  {author} {\bibfnamefont {W.}~\bibnamefont {Ratcliff}}, \bibinfo {author}
  {\bibfnamefont {J.~W.}\ \bibnamefont {Lynn}}, \bibinfo {author}
  {\bibfnamefont {S.}~\bibnamefont {Park}}, \ and\ \bibinfo {author}
  {\bibfnamefont {S.-W.}\ \bibnamefont {Cheong}},\ }\href@noop {} {\bibfield
  {journal} {\bibinfo  {journal} {Phys. Rev. B}\ }\textbf {\bibinfo {volume}
  {73}},\ \bibinfo {pages} {100406(R)} (\bibinfo {year}
  {2006}{\natexlab{a}})}\BibitemShut {NoStop}%
\bibitem [{\citenamefont {Hur}\ \emph {et~al.}(2004{\natexlab{b}})\citenamefont
  {Hur}, \citenamefont {Park}, \citenamefont {Sharma}, \citenamefont {Ahn},
  \citenamefont {Guha},\ and\ \citenamefont {Cheong}}]{hur-nat}%
  \BibitemOpen
  \bibfield  {author} {\bibinfo {author} {\bibfnamefont {N.}~\bibnamefont
  {Hur}}, \bibinfo {author} {\bibfnamefont {S.}~\bibnamefont {Park}}, \bibinfo
  {author} {\bibfnamefont {P.~A.}\ \bibnamefont {Sharma}}, \bibinfo {author}
  {\bibfnamefont {J.~S.}\ \bibnamefont {Ahn}}, \bibinfo {author} {\bibfnamefont
  {S.}~\bibnamefont {Guha}}, \ and\ \bibinfo {author} {\bibfnamefont {S.-W.}\
  \bibnamefont {Cheong}},\ }\href@noop {} {\bibfield  {journal} {\bibinfo
  {journal} {Nature}\ }\textbf {\bibinfo {volume} {429}},\ \bibinfo {pages}
  {392} (\bibinfo {year} {2004}{\natexlab{b}})}\BibitemShut {NoStop}%
\bibitem [{\citenamefont {Inomata}\ and\ \citenamefont
  {Kohn}(1996)}]{inomata-rmojpcm}%
  \BibitemOpen
  \bibfield  {author} {\bibinfo {author} {\bibfnamefont {A.}~\bibnamefont
  {Inomata}}\ and\ \bibinfo {author} {\bibfnamefont {K.}~\bibnamefont {Kohn}},\
  }\href@noop {} {\bibfield  {journal} {\bibinfo  {journal} {J. Phys.: Condens.
  Matter}\ }\textbf {\bibinfo {volume} {8}},\ \bibinfo {pages} {2673} (\bibinfo
  {year} {1996})}\BibitemShut {NoStop}%
\bibitem [{\citenamefont {Nakamura}\ and\ \citenamefont
  {Kohn}(1997)}]{nakamur-ferro}%
  \BibitemOpen
  \bibfield  {author} {\bibinfo {author} {\bibfnamefont {N.}~\bibnamefont
  {Nakamura}}\ and\ \bibinfo {author} {\bibfnamefont {K.}~\bibnamefont
  {Kohn}},\ }\href@noop {} {\bibfield  {journal} {\bibinfo  {journal}
  {Ferroelectrics}\ }\textbf {\bibinfo {volume} {204}},\ \bibinfo {pages} {107}
  (\bibinfo {year} {1997})}\BibitemShut {NoStop}%
\bibitem [{\citenamefont {Chattopadhyay}\ \emph {et~al.}(2016)\citenamefont
  {Chattopadhyay}, \citenamefont {Bal´edent}, \citenamefont {Damay},
  \citenamefont {Gukasov}, \citenamefont {Moshopoulou}, \citenamefont
  {Auban-Senzier}, \citenamefont {Pasquier}, \citenamefont {Andr´e},
  \citenamefont {Porcher}, \citenamefont {Elkaim}, \citenamefont {Doubrovsky},
  \citenamefont {Greenblatt},\ and\ \citenamefont
  {Foury-Leylekian}}]{sumanta-nmo}%
  \BibitemOpen
  \bibfield  {author} {\bibinfo {author} {\bibfnamefont {S.}~\bibnamefont
  {Chattopadhyay}}, \bibinfo {author} {\bibfnamefont {V.}~\bibnamefont
  {Bal´edent}}, \bibinfo {author} {\bibfnamefont {F.}~\bibnamefont {Damay}},
  \bibinfo {author} {\bibfnamefont {A.}~\bibnamefont {Gukasov}}, \bibinfo
  {author} {\bibfnamefont {E.}~\bibnamefont {Moshopoulou}}, \bibinfo {author}
  {\bibfnamefont {P.}~\bibnamefont {Auban-Senzier}}, \bibinfo {author}
  {\bibfnamefont {C.}~\bibnamefont {Pasquier}}, \bibinfo {author}
  {\bibfnamefont {G.}~\bibnamefont {Andr´e}}, \bibinfo {author} {\bibfnamefont
  {F.}~\bibnamefont {Porcher}}, \bibinfo {author} {\bibfnamefont
  {E.}~\bibnamefont {Elkaim}}, \bibinfo {author} {\bibfnamefont
  {C.}~\bibnamefont {Doubrovsky}}, \bibinfo {author} {\bibfnamefont
  {M.}~\bibnamefont {Greenblatt}}, \ and\ \bibinfo {author} {\bibfnamefont
  {P.}~\bibnamefont {Foury-Leylekian}},\ }\href@noop {} {\bibfield  {journal}
  {\bibinfo  {journal} {Phys. Rev. B}\ }\textbf {\bibinfo {volume} {93}},\
  \bibinfo {pages} {104406} (\bibinfo {year} {2016})}\BibitemShut {NoStop}%
\bibitem [{\citenamefont {Yang}\ \emph {et~al.}(2016)\citenamefont {Yang},
  \citenamefont {Li}, \citenamefont {Liu}, \citenamefont {Li}, \citenamefont
  {Yan}, \citenamefont {Zeng}, \citenamefont {Qin}, \citenamefont {Gao},\ and\
  \citenamefont {Liu}}]{yang-sr}%
  \BibitemOpen
  \bibfield  {author} {\bibinfo {author} {\bibfnamefont {L.}~\bibnamefont
  {Yang}}, \bibinfo {author} {\bibfnamefont {X.}~\bibnamefont {Li}}, \bibinfo
  {author} {\bibfnamefont {M.~F.}\ \bibnamefont {Liu}}, \bibinfo {author}
  {\bibfnamefont {P.~L.}\ \bibnamefont {Li}}, \bibinfo {author} {\bibfnamefont
  {Z.~B.}\ \bibnamefont {Yan}}, \bibinfo {author} {\bibfnamefont
  {M.}~\bibnamefont {Zeng}}, \bibinfo {author} {\bibfnamefont {M.~H.}\
  \bibnamefont {Qin}}, \bibinfo {author} {\bibfnamefont {X.~S.}\ \bibnamefont
  {Gao}}, \ and\ \bibinfo {author} {\bibfnamefont {J.-M.}\ \bibnamefont
  {Liu}},\ }\href@noop {} {\bibfield  {journal} {\bibinfo  {journal} {Sci.
  Rep.}\ }\textbf {\bibinfo {volume} {6}},\ \bibinfo {pages} {34767} (\bibinfo
  {year} {2016})}\BibitemShut {NoStop}%
\bibitem [{\citenamefont {Sanina}\ \emph {et~al.}(2011)\citenamefont {Sanina},
  \citenamefont {Golovenchits}, \citenamefont {Zalesskii},\ and\ \citenamefont
  {Scheglov}}]{sanina-jpcm}%
  \BibitemOpen
  \bibfield  {author} {\bibinfo {author} {\bibfnamefont {V.~A.}\ \bibnamefont
  {Sanina}}, \bibinfo {author} {\bibfnamefont {E.~I.}\ \bibnamefont
  {Golovenchits}}, \bibinfo {author} {\bibfnamefont {V.~G.}\ \bibnamefont
  {Zalesskii}}, \ and\ \bibinfo {author} {\bibfnamefont {M.~P.}\ \bibnamefont
  {Scheglov}},\ }\href@noop {} {\bibfield  {journal} {\bibinfo  {journal} {J.
  Phys.: Condens. Matter}\ }\textbf {\bibinfo {volume} {23}},\ \bibinfo {pages}
  {456003} (\bibinfo {year} {2011})}\BibitemShut {NoStop}%
\bibitem [{\citenamefont {Kagomiya}\ \emph {et~al.}(2002)\citenamefont
  {Kagomiya}, \citenamefont {Kohn},\ and\ \citenamefont
  {Uchiyama}}]{isao-ferro}%
  \BibitemOpen
  \bibfield  {author} {\bibinfo {author} {\bibfnamefont {I.}~\bibnamefont
  {Kagomiya}}, \bibinfo {author} {\bibfnamefont {K.}~\bibnamefont {Kohn}}, \
  and\ \bibinfo {author} {\bibfnamefont {T.}~\bibnamefont {Uchiyama}},\
  }\href@noop {} {\bibfield  {journal} {\bibinfo  {journal} {Ferroelectrics}\
  }\textbf {\bibinfo {volume} {280}},\ \bibinfo {pages} {131} (\bibinfo {year}
  {2002})}\BibitemShut {NoStop}%
\bibitem [{\citenamefont {Cheong}(2007)}]{cheong-nm}%
  \BibitemOpen
  \bibfield  {author} {\bibinfo {author} {\bibfnamefont {M.}~\bibnamefont
  {Cheong}, \bibfnamefont {S.~W.and~Mostovoy}},\ }\href@noop {} {\bibfield
  {journal} {\bibinfo  {journal} {Nat. Mater.}\ }\textbf {\bibinfo {volume}
  {6}},\ \bibinfo {pages} {13} (\bibinfo {year} {2007})}\BibitemShut {NoStop}%
\bibitem [{\citenamefont {Brink}\ and\ \citenamefont
  {Khomskii}(2008)}]{Jeroen-exstric}%
  \BibitemOpen
  \bibfield  {author} {\bibinfo {author} {\bibfnamefont {J.~v.~d.}\
  \bibnamefont {Brink}}\ and\ \bibinfo {author} {\bibfnamefont {D.~I.}\
  \bibnamefont {Khomskii}},\ }\href@noop {} {\bibfield  {journal} {\bibinfo
  {journal} {J. Phys.: Condens. Matter}\ }\textbf {\bibinfo {volume} {20}},\
  \bibinfo {pages} {434217} (\bibinfo {year} {2008})}\BibitemShut {NoStop}%
\bibitem [{\citenamefont {Balédent}\ \emph {et~al.}(2015)\citenamefont
  {Balédent}, \citenamefont {Chattopadhyay}, \citenamefont {Fertey},
  \citenamefont {Lepetit}, \citenamefont {Greenblatt}, \citenamefont {Wanklyn},
  \citenamefont {Saouma}, \citenamefont {Jang},\ and\ \citenamefont
  {Foury-Leylekian}}]{sumanta-rmoprl}%
  \BibitemOpen
  \bibfield  {author} {\bibinfo {author} {\bibfnamefont {V.}~\bibnamefont
  {Balédent}}, \bibinfo {author} {\bibfnamefont {S.}~\bibnamefont
  {Chattopadhyay}}, \bibinfo {author} {\bibfnamefont {P.}~\bibnamefont
  {Fertey}}, \bibinfo {author} {\bibfnamefont {M.~B.}\ \bibnamefont {Lepetit}},
  \bibinfo {author} {\bibfnamefont {M.}~\bibnamefont {Greenblatt}}, \bibinfo
  {author} {\bibfnamefont {B.}~\bibnamefont {Wanklyn}}, \bibinfo {author}
  {\bibfnamefont {F.~O.}\ \bibnamefont {Saouma}}, \bibinfo {author}
  {\bibfnamefont {J.~I.}\ \bibnamefont {Jang}}, \ and\ \bibinfo {author}
  {\bibfnamefont {P.}~\bibnamefont {Foury-Leylekian}},\ }\href@noop {}
  {\bibfield  {journal} {\bibinfo  {journal} {Phys. Rev. Lett.}\ }\textbf
  {\bibinfo {volume} {114}},\ \bibinfo {pages} {117601} (\bibinfo {year}
  {2015})}\BibitemShut {NoStop}%
\bibitem [{\citenamefont {Noda}\ \emph {et~al.}(2008)\citenamefont {Noda},
  \citenamefont {Kimura}, \citenamefont {Fukunaga}, \citenamefont {Kobayashi},
  \citenamefont {Kagomiya},\ and\ \citenamefont {Kohn}}]{Noda-rmo}%
  \BibitemOpen
  \bibfield  {author} {\bibinfo {author} {\bibfnamefont {Y.}~\bibnamefont
  {Noda}}, \bibinfo {author} {\bibfnamefont {H.}~\bibnamefont {Kimura}},
  \bibinfo {author} {\bibfnamefont {M.}~\bibnamefont {Fukunaga}}, \bibinfo
  {author} {\bibfnamefont {S.}~\bibnamefont {Kobayashi}}, \bibinfo {author}
  {\bibfnamefont {I.}~\bibnamefont {Kagomiya}}, \ and\ \bibinfo {author}
  {\bibfnamefont {K.}~\bibnamefont {Kohn}},\ }\href@noop {} {\bibfield
  {journal} {\bibinfo  {journal} {J. Phys.: Condens. Matter}\ }\textbf
  {\bibinfo {volume} {20}},\ \bibinfo {pages} {434206} (\bibinfo {year}
  {2008})}\BibitemShut {NoStop}%
\bibitem [{\citenamefont {Radaelli}\ and\ \citenamefont
  {Chapon}(2008)}]{radaelli-jpcm}%
  \BibitemOpen
  \bibfield  {author} {\bibinfo {author} {\bibfnamefont {P.~G.}\ \bibnamefont
  {Radaelli}}\ and\ \bibinfo {author} {\bibfnamefont {L.~C.}\ \bibnamefont
  {Chapon}},\ }\href@noop {} {\bibfield  {journal} {\bibinfo  {journal} {J.
  Phys.: Condens. Matter}\ }\textbf {\bibinfo {volume} {20}},\ \bibinfo {pages}
  {434213} (\bibinfo {year} {2008})}\BibitemShut {NoStop}%
\bibitem [{\citenamefont {Chattopadhyay}\ \emph {et~al.}(2017)\citenamefont
  {Chattopadhyay}, \citenamefont {Petit}, \citenamefont {Ressouche},
  \citenamefont {Raymond}, \citenamefont {Balédent}, \citenamefont {Yahia},
  \citenamefont {Peng}, \citenamefont {Robert}, \citenamefont {Lepetit},
  \citenamefont {Greenblatt},\ and\ \citenamefont
  {Foury-Leylekian}}]{sumanta-sr}%
  \BibitemOpen
  \bibfield  {author} {\bibinfo {author} {\bibfnamefont {S.}~\bibnamefont
  {Chattopadhyay}}, \bibinfo {author} {\bibfnamefont {S.}~\bibnamefont
  {Petit}}, \bibinfo {author} {\bibfnamefont {E.}~\bibnamefont {Ressouche}},
  \bibinfo {author} {\bibfnamefont {S.}~\bibnamefont {Raymond}}, \bibinfo
  {author} {\bibfnamefont {V.}~\bibnamefont {Balédent}}, \bibinfo {author}
  {\bibfnamefont {G.}~\bibnamefont {Yahia}}, \bibinfo {author} {\bibfnamefont
  {W.}~\bibnamefont {Peng}}, \bibinfo {author} {\bibfnamefont {J.}~\bibnamefont
  {Robert}}, \bibinfo {author} {\bibfnamefont {M.-B.}\ \bibnamefont {Lepetit}},
  \bibinfo {author} {\bibfnamefont {M.}~\bibnamefont {Greenblatt}}, \ and\
  \bibinfo {author} {\bibfnamefont {P.}~\bibnamefont {Foury-Leylekian}},\
  }\href@noop {} {\bibfield  {journal} {\bibinfo  {journal} {Sci. Rep.}\
  }\textbf {\bibinfo {volume} {7}},\ \bibinfo {pages} {14506} (\bibinfo {year}
  {2017})}\BibitemShut {NoStop}%
\bibitem [{\citenamefont {Blake}\ \emph {et~al.}(2005)\citenamefont {Blake},
  \citenamefont {Chapon}, \citenamefont {Radaelli}, \citenamefont {Park},
  \citenamefont {Hur}, \citenamefont {Cheong},\ and\ \citenamefont
  {Rodríguez-Carvajal}}]{Blake-prb}%
  \BibitemOpen
  \bibfield  {author} {\bibinfo {author} {\bibfnamefont {G.~R.}\ \bibnamefont
  {Blake}}, \bibinfo {author} {\bibfnamefont {L.~C.}\ \bibnamefont {Chapon}},
  \bibinfo {author} {\bibfnamefont {P.~G.}\ \bibnamefont {Radaelli}}, \bibinfo
  {author} {\bibfnamefont {S.}~\bibnamefont {Park}}, \bibinfo {author}
  {\bibfnamefont {N.}~\bibnamefont {Hur}}, \bibinfo {author} {\bibfnamefont
  {S.-W.}\ \bibnamefont {Cheong}}, \ and\ \bibinfo {author} {\bibfnamefont
  {J.}~\bibnamefont {Rodríguez-Carvajal}},\ }\href@noop {} {\bibfield
  {journal} {\bibinfo  {journal} {Phys. Rev. B}\ }\textbf {\bibinfo {volume}
  {71}},\ \bibinfo {pages} {214402} (\bibinfo {year} {2005})}\BibitemShut
  {NoStop}%
\bibitem [{\citenamefont {Zhao}\ \emph
  {et~al.}(2014{\natexlab{a}})\citenamefont {Zhao}, \citenamefont {Liu},
  \citenamefont {Li}, \citenamefont {Lin}, \citenamefont {Yan}, \citenamefont
  {Dong},\ and\ \citenamefont {Liu}}]{Zhao-sr}%
  \BibitemOpen
  \bibfield  {author} {\bibinfo {author} {\bibfnamefont {Z.~Y.}\ \bibnamefont
  {Zhao}}, \bibinfo {author} {\bibfnamefont {M.~F.}\ \bibnamefont {Liu}},
  \bibinfo {author} {\bibfnamefont {X.}~\bibnamefont {Li}}, \bibinfo {author}
  {\bibfnamefont {L.}~\bibnamefont {Lin}}, \bibinfo {author} {\bibfnamefont
  {Z.~B.}\ \bibnamefont {Yan}}, \bibinfo {author} {\bibfnamefont
  {S.}~\bibnamefont {Dong}}, \ and\ \bibinfo {author} {\bibfnamefont {J.~M.}\
  \bibnamefont {Liu}},\ }\href@noop {} {\bibfield  {journal} {\bibinfo
  {journal} {Sci. Rep.}\ }\textbf {\bibinfo {volume} {4}},\ \bibinfo {pages}
  {3984} (\bibinfo {year} {2014}{\natexlab{a}})}\BibitemShut {NoStop}%
\bibitem [{\citenamefont {Ratcliff~II}\ \emph {et~al.}(2005)\citenamefont
  {Ratcliff~II}, \citenamefont {Kiryukhin}, \citenamefont {Kenzelmann},
  \citenamefont {Lee}, \citenamefont {Erwin}, \citenamefont {Schefer},
  \citenamefont {Hur}, \citenamefont {Park},\ and\ \citenamefont
  {Cheong}}]{Ratcliff-prb}%
  \BibitemOpen
  \bibfield  {author} {\bibinfo {author} {\bibfnamefont {W.}~\bibnamefont
  {Ratcliff~II}}, \bibinfo {author} {\bibfnamefont {V.}~\bibnamefont
  {Kiryukhin}}, \bibinfo {author} {\bibfnamefont {M.}~\bibnamefont
  {Kenzelmann}}, \bibinfo {author} {\bibfnamefont {S.-H.}\ \bibnamefont {Lee}},
  \bibinfo {author} {\bibfnamefont {R.}~\bibnamefont {Erwin}}, \bibinfo
  {author} {\bibfnamefont {J.}~\bibnamefont {Schefer}}, \bibinfo {author}
  {\bibfnamefont {N.}~\bibnamefont {Hur}}, \bibinfo {author} {\bibfnamefont
  {S.}~\bibnamefont {Park}}, \ and\ \bibinfo {author} {\bibfnamefont {S.-W.}\
  \bibnamefont {Cheong}},\ }\href@noop {} {\bibfield  {journal} {\bibinfo
  {journal} {Phys. Rev. B}\ }\textbf {\bibinfo {volume} {72}},\ \bibinfo
  {pages} {060407(R)} (\bibinfo {year} {2005})}\BibitemShut {NoStop}%
\bibitem [{\citenamefont {Sushkov}\ \emph {et~al.}(2014)\citenamefont
  {Sushkov}, \citenamefont {Kant}, \citenamefont {Schiebl}, \citenamefont
  {Shuvaev}, \citenamefont {Pimenov}, \citenamefont {Pimenov}, \citenamefont
  {Lorenz}, \citenamefont {Park}, \citenamefont {Cheong}, \citenamefont
  {Mostovoy},\ and\ \citenamefont {Drew}}]{Sushkov-prb}%
  \BibitemOpen
  \bibfield  {author} {\bibinfo {author} {\bibfnamefont {A.~B.}\ \bibnamefont
  {Sushkov}}, \bibinfo {author} {\bibfnamefont {C.}~\bibnamefont {Kant}},
  \bibinfo {author} {\bibfnamefont {M.}~\bibnamefont {Schiebl}}, \bibinfo
  {author} {\bibfnamefont {A.~M.}\ \bibnamefont {Shuvaev}}, \bibinfo {author}
  {\bibfnamefont {A.}~\bibnamefont {Pimenov}}, \bibinfo {author} {\bibfnamefont
  {A.}~\bibnamefont {Pimenov}}, \bibinfo {author} {\bibfnamefont
  {B.}~\bibnamefont {Lorenz}}, \bibinfo {author} {\bibfnamefont
  {S.}~\bibnamefont {Park}}, \bibinfo {author} {\bibfnamefont {S.-W.}\
  \bibnamefont {Cheong}}, \bibinfo {author} {\bibfnamefont {M.}~\bibnamefont
  {Mostovoy}}, \ and\ \bibinfo {author} {\bibfnamefont {H.~D.}\ \bibnamefont
  {Drew}},\ }\href@noop {} {\bibfield  {journal} {\bibinfo  {journal} {Phys.
  Rev. B}\ }\textbf {\bibinfo {volume} {90}},\ \bibinfo {pages} {054417}
  (\bibinfo {year} {2014})}\BibitemShut {NoStop}%
\bibitem [{\citenamefont {Lutterotti}\ \emph {et~al.}(1997)\citenamefont
  {Lutterotti}, \citenamefont {Matthies}, \citenamefont {Wenk}, \citenamefont
  {Schultz},\ and\ \citenamefont {Richardson}}]{maud-jap}%
  \BibitemOpen
  \bibfield  {author} {\bibinfo {author} {\bibfnamefont {L.}~\bibnamefont
  {Lutterotti}}, \bibinfo {author} {\bibfnamefont {S.}~\bibnamefont
  {Matthies}}, \bibinfo {author} {\bibfnamefont {H.-R.}\ \bibnamefont {Wenk}},
  \bibinfo {author} {\bibfnamefont {A.~S.}\ \bibnamefont {Schultz}}, \ and\
  \bibinfo {author} {\bibfnamefont {J.~W.}\ \bibnamefont {Richardson}},\
  }\href@noop {} {\bibfield  {journal} {\bibinfo  {journal} {J. Appl. Phys.}\
  }\textbf {\bibinfo {volume} {81}},\ \bibinfo {pages} {594} (\bibinfo {year}
  {1997})}\BibitemShut {NoStop}%
\bibitem [{\citenamefont {Mihailova}\ \emph {et~al.}(2005)\citenamefont
  {Mihailova}, \citenamefont {Gospodinov}, \citenamefont {Güttler},
  \citenamefont {Yen}, \citenamefont {Litvinchuk},\ and\ \citenamefont
  {Iliev}}]{Mihailova-rmo}%
  \BibitemOpen
  \bibfield  {author} {\bibinfo {author} {\bibfnamefont {B.}~\bibnamefont
  {Mihailova}}, \bibinfo {author} {\bibfnamefont {M.~M.}\ \bibnamefont
  {Gospodinov}}, \bibinfo {author} {\bibfnamefont {B.}~\bibnamefont
  {Güttler}}, \bibinfo {author} {\bibfnamefont {F.}~\bibnamefont {Yen}},
  \bibinfo {author} {\bibfnamefont {A.~P.}\ \bibnamefont {Litvinchuk}}, \ and\
  \bibinfo {author} {\bibfnamefont {M.~N.}\ \bibnamefont {Iliev}},\ }\href@noop
  {} {\bibfield  {journal} {\bibinfo  {journal} {Phys. Rev. B}\ }\textbf
  {\bibinfo {volume} {71}},\ \bibinfo {pages} {172301} (\bibinfo {year}
  {2005})}\BibitemShut {NoStop}%
\bibitem [{\citenamefont {Cruz}\ \emph
  {et~al.}(2006{\natexlab{b}})\citenamefont {Cruz}, \citenamefont {Yen},
  \citenamefont {Lorenz}, \citenamefont {Park}, \citenamefont {Cheong},
  \citenamefont {Gospodinov}, \citenamefont {Ratcliff}, \citenamefont {Lynn},\
  and\ \citenamefont {Chu}}]{cruz-rmojap}%
  \BibitemOpen
  \bibfield  {author} {\bibinfo {author} {\bibfnamefont {C.~R.~d.}\
  \bibnamefont {Cruz}}, \bibinfo {author} {\bibfnamefont {F.}~\bibnamefont
  {Yen}}, \bibinfo {author} {\bibfnamefont {B.}~\bibnamefont {Lorenz}},
  \bibinfo {author} {\bibfnamefont {S.}~\bibnamefont {Park}}, \bibinfo {author}
  {\bibfnamefont {S.-W.}\ \bibnamefont {Cheong}}, \bibinfo {author}
  {\bibfnamefont {M.~M.}\ \bibnamefont {Gospodinov}}, \bibinfo {author}
  {\bibfnamefont {W.}~\bibnamefont {Ratcliff}}, \bibinfo {author}
  {\bibfnamefont {J.~W.}\ \bibnamefont {Lynn}}, \ and\ \bibinfo {author}
  {\bibfnamefont {C.~W.}\ \bibnamefont {Chu}},\ }\href@noop {} {\bibfield
  {journal} {\bibinfo  {journal} {J. Appl. Phys.}\ }\textbf {\bibinfo {volume}
  {99}},\ \bibinfo {pages} {08R103} (\bibinfo {year}
  {2006}{\natexlab{b}})}\BibitemShut {NoStop}%
\bibitem [{\citenamefont {Pecharsky}\ and\ \citenamefont
  {Gschneidner~Jr}(1999)}]{gs-mce}%
  \BibitemOpen
  \bibfield  {author} {\bibinfo {author} {\bibfnamefont {V.~K.}\ \bibnamefont
  {Pecharsky}}\ and\ \bibinfo {author} {\bibfnamefont {K.~A.}\ \bibnamefont
  {Gschneidner~Jr}},\ }\href@noop {} {\bibfield  {journal} {\bibinfo  {journal}
  {J. Mag. Mag Mater.}\ }\textbf {\bibinfo {volume} {200}},\ \bibinfo {pages}
  {44} (\bibinfo {year} {1999})}\BibitemShut {NoStop}%
\bibitem [{\citenamefont {Pathak}\ \emph {et~al.}(2010)\citenamefont {Pathak},
  \citenamefont {Dubenko}, \citenamefont {Karaca}, \citenamefont {Stadler},\
  and\ \citenamefont {Ali}}]{ali-rcp}%
  \BibitemOpen
  \bibfield  {author} {\bibinfo {author} {\bibfnamefont {A.~K.}\ \bibnamefont
  {Pathak}}, \bibinfo {author} {\bibfnamefont {I.}~\bibnamefont {Dubenko}},
  \bibinfo {author} {\bibfnamefont {H.~E.}\ \bibnamefont {Karaca}}, \bibinfo
  {author} {\bibfnamefont {S.}~\bibnamefont {Stadler}}, \ and\ \bibinfo
  {author} {\bibfnamefont {N.}~\bibnamefont {Ali}},\ }\href@noop {} {\bibfield
  {journal} {\bibinfo  {journal} {Appl. Phys. Lett.}\ }\textbf {\bibinfo
  {volume} {97}},\ \bibinfo {pages} {062505} (\bibinfo {year}
  {2010})}\BibitemShut {NoStop}%
\bibitem [{\citenamefont {Hernando}\ \emph {et~al.}(2009)\citenamefont
  {Hernando}, \citenamefont {Sánchez~Llamazares}, \citenamefont {Prida},
  \citenamefont {Baldomir}, \citenamefont {Serantes}, \citenamefont {Ilyn},\
  and\ \citenamefont {González}}]{hernado-rcp}%
  \BibitemOpen
  \bibfield  {author} {\bibinfo {author} {\bibfnamefont {B.}~\bibnamefont
  {Hernando}}, \bibinfo {author} {\bibfnamefont {J.~L.}\ \bibnamefont
  {Sánchez~Llamazares}}, \bibinfo {author} {\bibfnamefont {V.~M.}\
  \bibnamefont {Prida}}, \bibinfo {author} {\bibfnamefont {D.}~\bibnamefont
  {Baldomir}}, \bibinfo {author} {\bibfnamefont {D.}~\bibnamefont {Serantes}},
  \bibinfo {author} {\bibfnamefont {M.}~\bibnamefont {Ilyn}}, \ and\ \bibinfo
  {author} {\bibfnamefont {J.}~\bibnamefont {González}},\ }\href@noop {}
  {\bibfield  {journal} {\bibinfo  {journal} {Appl. Phys. Lett.}\ }\textbf
  {\bibinfo {volume} {94}},\ \bibinfo {pages} {222502} (\bibinfo {year}
  {2009})}\BibitemShut {NoStop}%
\bibitem [{\citenamefont {Phan}\ and\ \citenamefont {Yu}(2007)}]{phan-jmmm}%
  \BibitemOpen
  \bibfield  {author} {\bibinfo {author} {\bibfnamefont {M.-H.}\ \bibnamefont
  {Phan}}\ and\ \bibinfo {author} {\bibfnamefont {S.-C.}\ \bibnamefont {Yu}},\
  }\href@noop {} {\bibfield  {journal} {\bibinfo  {journal} {J. Mag. Mag
  Mater.}\ }\textbf {\bibinfo {volume} {308}},\ \bibinfo {pages} {325}
  (\bibinfo {year} {2007})}\BibitemShut {NoStop}%
\bibitem [{\citenamefont {Zhao}\ \emph
  {et~al.}(2014{\natexlab{b}})\citenamefont {Zhao}, \citenamefont {Liu},
  \citenamefont {Li}, \citenamefont {Wang}, \citenamefont {Yan}, \citenamefont
  {Wang},\ and\ \citenamefont {Liu}}]{Zhao-Al-doped}%
  \BibitemOpen
  \bibfield  {author} {\bibinfo {author} {\bibfnamefont {Z.~Y.}\ \bibnamefont
  {Zhao}}, \bibinfo {author} {\bibfnamefont {M.~F.}\ \bibnamefont {Liu}},
  \bibinfo {author} {\bibfnamefont {X.}~\bibnamefont {Li}}, \bibinfo {author}
  {\bibfnamefont {J.~X.}\ \bibnamefont {Wang}}, \bibinfo {author}
  {\bibfnamefont {Z.~B.}\ \bibnamefont {Yan}}, \bibinfo {author} {\bibfnamefont
  {K.~F.}\ \bibnamefont {Wang}}, \ and\ \bibinfo {author} {\bibfnamefont
  {J.~M.}\ \bibnamefont {Liu}},\ }\href@noop {} {\bibfield  {journal} {\bibinfo
   {journal} {J. Appl. Phys.}\ }\textbf {\bibinfo {volume} {116}},\ \bibinfo
  {pages} {054104} (\bibinfo {year} {2014}{\natexlab{b}})}\BibitemShut
  {NoStop}%
\bibitem [{\citenamefont {Zhao}\ \emph {et~al.}(2015)\citenamefont {Zhao},
  \citenamefont {Wang}, \citenamefont {Lin}, \citenamefont {Liu}, \citenamefont
  {Li}, \citenamefont {Yan},\ and\ \citenamefont {Liu}}]{Zhao-Y-doped}%
  \BibitemOpen
  \bibfield  {author} {\bibinfo {author} {\bibfnamefont {Z.~Y.}\ \bibnamefont
  {Zhao}}, \bibinfo {author} {\bibfnamefont {Y.~L.}\ \bibnamefont {Wang}},
  \bibinfo {author} {\bibfnamefont {L.}~\bibnamefont {Lin}}, \bibinfo {author}
  {\bibfnamefont {M.~F.}\ \bibnamefont {Liu}}, \bibinfo {author} {\bibfnamefont
  {X.}~\bibnamefont {Li}}, \bibinfo {author} {\bibfnamefont {Z.~B.}\
  \bibnamefont {Yan}}, \ and\ \bibinfo {author} {\bibfnamefont {J.~M.}\
  \bibnamefont {Liu}},\ }\href@noop {} {\bibfield  {journal} {\bibinfo
  {journal} {J. Appl. Phys.}\ }\textbf {\bibinfo {volume} {118}},\ \bibinfo
  {pages} {174105} (\bibinfo {year} {2015})}\BibitemShut {NoStop}%
\end{thebibliography}
%

\end{document}